\documentclass[9pt,arxiv]{lapreprint}

\usepackage[version=4]{mhchem} 
\usepackage{siunitx}    
\usepackage{pdflscape}  
\usepackage{rotating}   
\usepackage{textgreek}  
\usepackage{gensymb}    
\usepackage[misc]{ifsym} 
\usepackage{orcidlink}  
\usepackage{listings}   
\usepackage{colortbl}   
\usepackage{tabularx}   
\usepackage{longtable}  
\usepackage{subcaption}
\usepackage{multirow}
\usepackage{snotez}     
\usepackage{csquotes}   
\usepackage{color,soul}	

\DeclareSIUnit\Molar{M}

\usepackage[			
	backend=bibtex,      
    style=authoryear,   
	natbib=true,		
	isbn=false,
	url=false,
	eprint=false,
	hyperref=true,	    
	alldates=year,      
]{biblatex}

\title{Compressive Raman microspectroscopy parallelized by single-photon avalanche diode arrays}

\author[1 \orcidlink{0000-0002-0766-6564} \authfn{1}]{Clémence Gentner}
\author[2 \authfn{1}]{Samuel Burri}
\author[2 \orcidlink{0000-0002-0620-3365}]{Edoardo Charbon}
\author[2 \orcidlink{0000-0002-6636-6596} \Letter]{Claudio Bruschini}
\author[1 \orcidlink{0000-0002-2426-0371} \Letter]{Hilton B. de Aguiar}

\affil[1]{Laboratoire Kastler Brossel, ENS-Université PSL, CNRS, Sorbonne Université, Collège de France, 24 rue Lhomond, 75005 Paris, France}
\affil[2]{Advanced Quantum Architecture Lab (AQUA), EPFL, 2002 Neuchâtel, Switzerland}

\metadata[]{\Letter\hspace{.5ex} For correspondence}{claudio.bruschini@epfl.ch, h.aguiar@lkb.ens.fr \\ \url{www.lkb.upmc.fr/opticalimaging/}\\ \url{www.epfl.ch/labs/aqua/}}
\metadata[\authfn{1}]{}{These authors contributed equally to the work.}
\metadata[]{Funding}{This work was supported by the Emergence program from Sorbonne University (FANCIER).}

\leadauthor{Gentner}
\shorttitle{Compressive Raman with SPAD arrays}

\begin{document}
\maketitle
\begin{abstract}
We demonstrate an efficient and scalable compressive Raman parallelization scheme based on single-photon avalanche diode (SPAD) arrays to reach pixel dwell times of 23 µs, representing over 10$\times$ speed-up using the otherwise weak spontaneous Raman effect.
\end{abstract}

Compressive Raman has enabled high-speed spectral imaging by compressing the hyperspectrum at its acquisition stage, exploiting data sparsity \parencite{Wilcox2013,Cebeci2018,Lin2022}. The hardware of this emerging technology relies on a programmable spectrometer based on a spatial light modulator coupled to a single-pixel detector (instead of the more costly and read-out noise limited charge-coupled devices). While acquisition and processing times are already considerably reduced compared to conventional Raman imaging, further speed gains could in principle be achieved by increasing the excitation power (at constant energy). {Nevertheless, such an approach would be fundamentally limited by the detector's dead time. For instance, if one detects 10 photons}\parencite{Wilcox2013} {, current commercially available single-pixel SPADs would reach pixel dwell times} $\tau_{pdt}${ at best 1 }µs. Hence, spatial detection parallelization with high-throughput low-noise sensors constitutes the operating lever to reach real-time imaging.

While SPAD arrays are being developed at a fast pace \parencite{Bruschini2019}, they were until now incompatible with the requirements needed for high-speed imaging with the weak spontaneous Raman effect. Here, we introduce a novel compressive Raman spectrometer layout equipped with a new high-throughput SPAD linear array, LinoSPAD2 \parencite{Lubin2021}, to heavily parallelize the single-pixel approach. In addition to data compression in the spectral domain, this architecture allows spatial multiplexing with line scan illumination, leading to acquisition speeds only fundamentally limited by either inertial motion of 1D scanners or the detector's dead time. In this initial demonstration, an effective $\tau_{pdt}=\SI{23}{\micro\second}$ was achieved with the spontaneous Raman effect, with potential to reduce this twenty-fold, thus paving the way for video-rate imaging in spontaneous Raman microspectroscopy.

\begin{figure}[h]
    \centering
    \includegraphics[scale=0.44]{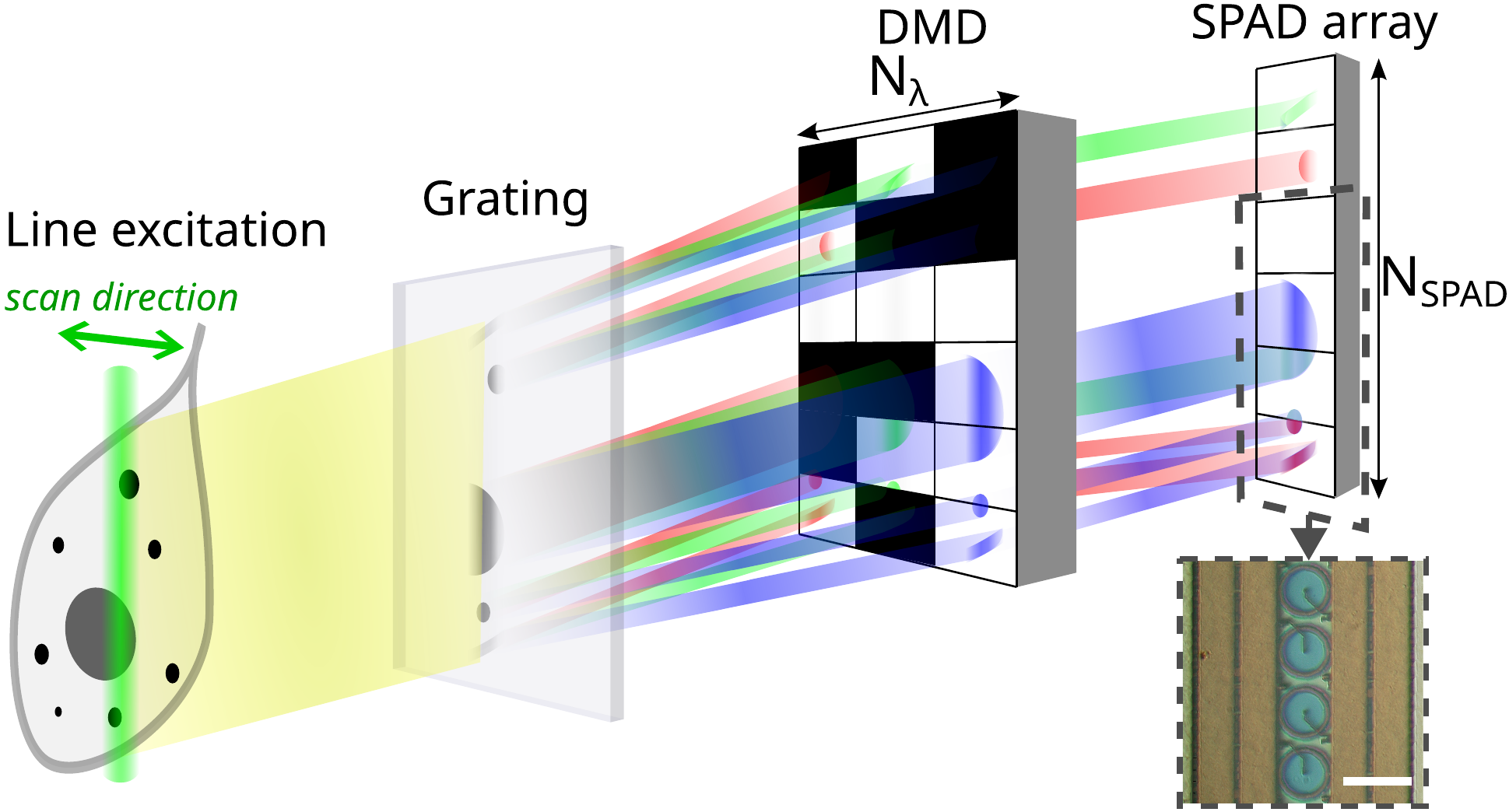}
    \caption{Parallelized compressive Raman microspectroscopy. The line-shaped laser light is focused on the sample and imaged onto the grating of the spectrometer, defining the $N_{\lambda}$ wavelength bins on the DMD. The dimension along the excitation line on the DMD parallelizes the detection of $N_{\text{SPAD}}$ spatial pixels. Black DMD pixels represent discarded wavelength bins. An image is formed by scanning  the line in its orthogonal direction. Inset: brightfield image of four pixels of the LinoSPAD2 array. Scale bar: 30 µm.}
    \label{fig:setup}
\end{figure}

In programmable spectrometers for compressive Raman, the emitted light from the point-scanned sample is steered into a conventional Czerny-Turner spectrometer\parencite{Wilcox2012,Wilcox2013,Sturm2019,Soldevila2019,Corden2018,Rehrauer2018,Scotte2019,Scotte2020}. The camera is replaced by a digital micromirror device (DMD) which sends specific spectral bins, selected according to the chemical identification algorithm used \parencite{Wilcox2012,Refregier2018,Soldevila2019}, towards a single-pixel SPAD. Spatial parallelization of this scheme (see Fig. \ref{fig:setup}) speeds up the acquisition by $N_{\text{SPAD}}\frac{PDE_{\text{array}}}{PDE_{\text{SP}}}$, where $N_{\text{SPAD}}$ is the number of spatial pixels and $PDE$ is the SPAD photon detection efficiency (either single-pixel (SP) or array). The LinoSPAD2 linear array is composed of 512 low-noise and high-sensitivity pixels implemented in 180~nm standard CMOS technology, with 1:1 direct read-out by a pair of custom FPGA boards \parencite{Burri2017}. This enables full reconfigurability of the photon counting and/or time-stamping functions.

\begin{figure}[h]
    \centering
    \includegraphics[scale=0.66]{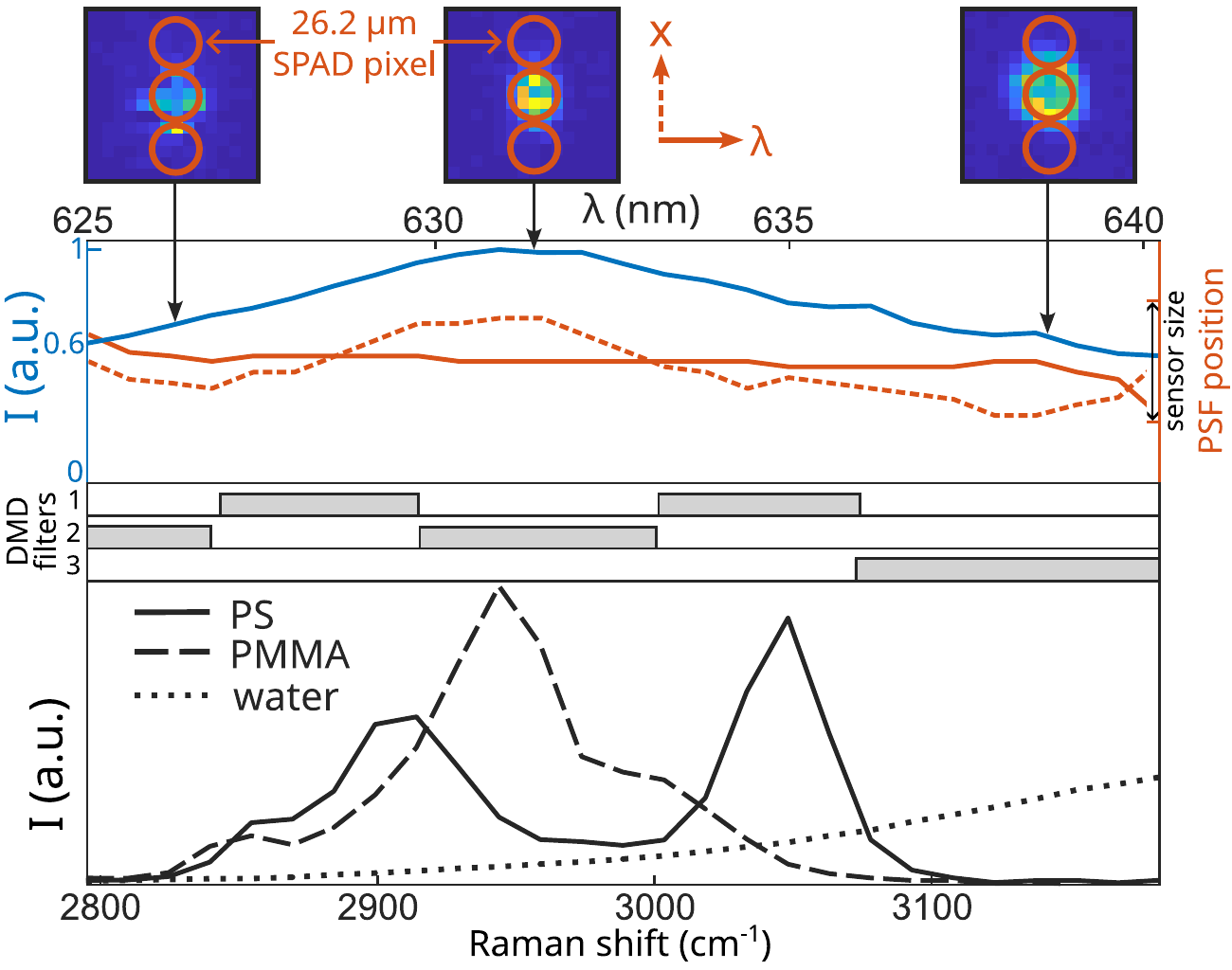}
    \caption{Characterization of the programmable spectrometer. (Top panel) Available spectral detection efficiency (blue) is limited by SPAD pixel crosstalk (see thumbnails for images acquired at the SPAD plane with an auxiliary camera). Displacement of the spot center (orange) in the $X$ (dashed) and $\lambda$ (continuous) directions. Optimal DMD filters (middle panel) for supervised identification of polystyrene (PS), polymethylmethacrylate (PMMA) and water spectra (bottom panel).}
    \label{fig:carac}
\end{figure}

One of the challenges in implementing this parallelization scheme is due to the DMD's own dispersion. Indeed, a DMD is a grating whose dispersion increases the sensitivity to alignment of the quality of the spectrometer PSF, giving rise to crosstalk between the SPAD pixels when moving away from the optimal wavelength (see Fig. \ref{fig:carac} thumbnails). This effect led to the widespread adoption of large-area single-pixel detectors (typically photomultipliers). Previously, we have described a methodology to overcome this issue and better control the dispersion in a programmable spectrometer using a \SI{190}{\micro\meter}-wide single-pixel SPAD \parencite{Sturm2019}. We now reach PSFs down to tens of µm, a typical size of SPADs in arrays, effectively enhancing the bandwidth coupling for a given spectral range, in this case the stretch resonances (C-H, O-H) spectral range (Fig. \ref{fig:carac}, bottom). Furthermore, the PSF undergoes a displacement mainly along the SPAD array direction and is stable enough over the spectral range, hence minimizing inter-pixel spectral coupling (Fig. \ref{fig:carac}, middle).

\begin{figure}
    \centering
    \includegraphics[scale=0.58]{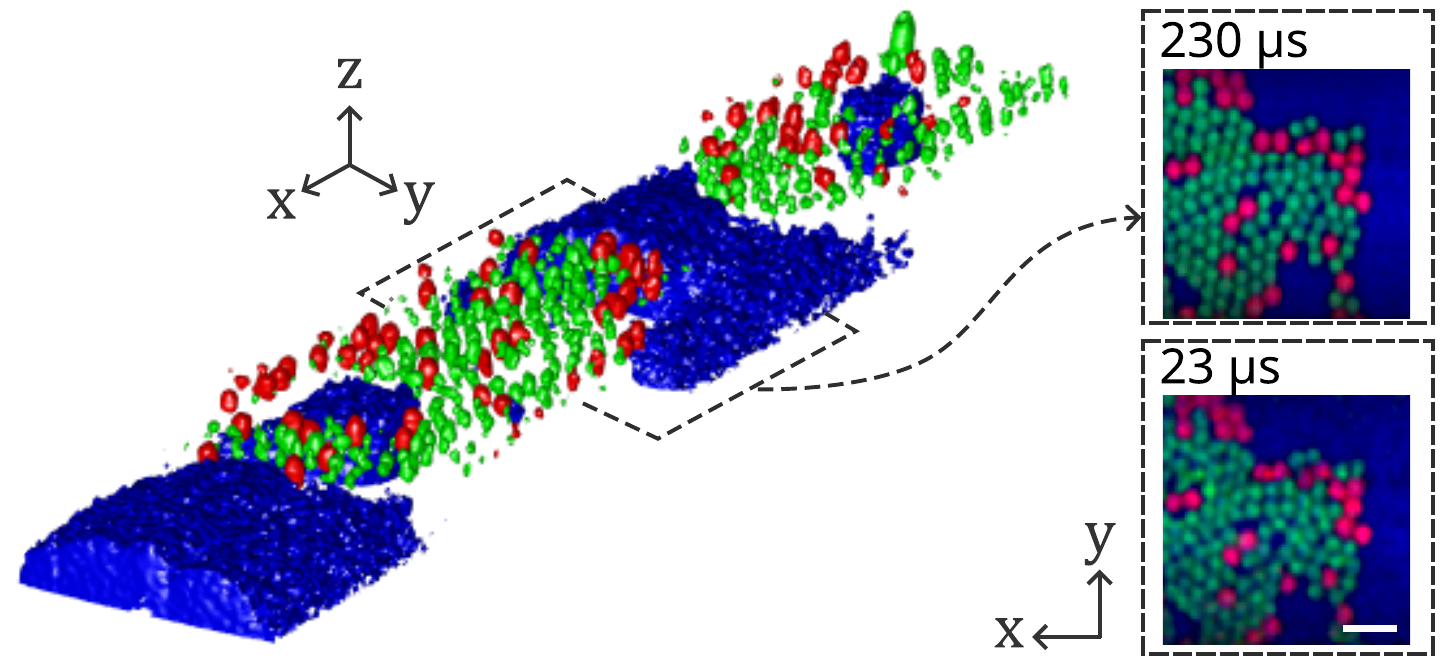}
    \caption{Compressive Raman image of beads made of PS (\SI{6}{\micro\meter}, red) and PMMA (\SI{5}{\micro\meter}, green) embedded in water (blue). The 3D rendering (left) is acquired at effective $\tau_{pdt}=\SI{230}{\micro\second}$. The two slices (right) correspond to 230 and \SI{23}{\micro\second}. Scale bar: \SI{20}{\micro\meter}}
    \label{fig:results}
\end{figure}

We demonstrate parallelized high-speed chemical imaging. Instead of 2D point scanning, the sample is excited by a laser line (Spectra Physics, Millennia eV10S, $\lambda$=532~\SI{}{\nano\meter}) that is 1D scanned (Thorlabs GVS012) using an objective (20$\times$, NA 0.5, Nikon) to focus the line to a length of 90~\SI{}{\micro\meter}. The epi-detected Raman signal from the line excitation is descanned and steered into the spectrometer coupled with the SPAD array through a 1D confocal layout using a 30~\SI{}{\micro\meter} slit at the entrance plane. The energy densities used were 4 and 40~\SI{}{\micro\joule\per\micro\meter\squared} for effective $\tau_{pdt}=\text{23}$ and 230~\SI{}{\micro\second}, respectively. We demonstrate the speed advantage using established supervised compressive Raman methods \parencite{Refregier2018} and as a sample, polystyrene (PS) and polymethylmethacrylate (PMMA) beads of 6 and 5~\SI{}{\micro\meter} diameters, respectively, on top of a coverslip covered with water (spectra shown in Fig. \ref{fig:carac}, bottom). Fig. \ref{fig:results} shows a 3D chemical image $(x,y,z)$ composed of 128$\times$531$\times$40 pixels. A standard denoising scheme was performed using Fourier domain Gaussian smoothing. Each $(x,y)$ slice (Fig. \ref{fig:results}, inset) was acquired by horizontally scanning at 10 and 1 ms line dwell time $\tau_{ldt}$, {orders of magnitude faster than conventional cameras in non-compressed approaches due to bottlenecks in data throughput}\parencite{Kumamoto2022}. \textit{Effective} $\tau_{pdt}$ is defined as $\tau_{ldt}\frac{N_{\text{DMD}}}{N_{\text{SPAD}}}$. The image acquired at effective $\tau_{pdt}=\SI{23}{\micro\second}$ ($\tau_{ldt}=1$ ms, $N_{\text{SPAD}}=128$) represents a $10\times$ improvement in comparison with previous approaches \parencite{Scotte2019,Soldevila2019}, despite significant losses in the layout (discussed below).

Further upgrades in the spectrometer and SPADs layout could increase this performance gain by at least $20\times$. We measured a  $PDE_{\text{array}}=7\%$, which can be increased to $20\%$ with light-coupling microlenses, therefore improving the detection efficiency by $3\times$. Another 4-fold gain will be provided by using all 512 SPAD array pixels instead of the current $N_{\text{SPAD}}=128$. An additional $2\times$ factor can be achieved by using both DMD outputs \parencite{Rehrauer2018}. We believe that further PSF engineering will enlarge the spectral bandwidth to cover also the fingerprint region, hence further improving spectral identification capabilities. Note that previous works have achieved $\tau_{pdt}=\SI{5}{\micro\second}$ for one filter \parencite{Rehrauer2018}, however reaching the fundamental limit of the system, whereas the parallelized scheme presented here achieves for one filter \SI{8}{\micro\second} with much space for improvement.

In conclusion, we demonstrate that an effective parallelization of the single-pixel-based compressive Raman microscope allows to considerably increase the image acquisition. These results pave the way to video-rate imaging with the inexpensive spontaneous Raman effect. Perspectives include label-free real-time biological or chemical identification solutions for applications such as histopathology and quality control.


\printbibliography

\if@endfloat\clearpage\processdelayedfloats\clearpage\fi

%
%
%


\end{document}